\documentclass [aps,twocolumn,prb,showpacs] {revtex4}
\usepackage{amsmath}
\usepackage{graphicx,epsfig,psfrag}
\usepackage{amssymb}

\renewcommand{\vec}[1]{{\bf \boldsymbol #1}}

\def\W{\Omega}

\def\a{\alpha}

\def\g{\gamma}
\def\d{\delta}

 \def\s{\sigma}

\begin{document}

\title{Crystalline phases in chiral ferromagnets: Destabilization of
  helical order}

\date{\today}

\author{Inga~Fischer$^1$, Nayana~Shah$^2$, and Achim~Rosch$^1$}
\affiliation{$^1$ Institute for Theoretical Physics, University of
  Cologne, 50937 Cologne, Germany, \\$^2$ Department of Physics,
  University of Illinois at
 Urbana-Champaign, 1110 W. Green St, Urbana, IL 61801, USA}
\



\begin{abstract}
  In chiral ferromagnets, weak spin-orbit interactions twists the
  ferromagnetic order into spirals leading to helical order. We
  investigate an extended Ginzburg-Landau theory of such systems where
  the helical order is destabilized in favour of crystalline phases.
  These crystalline phases are based on periodic arrangements of
  double-twist cylinders and are strongly reminiscent of blue phases
  in liquid crystals. We discuss the relevance of such blue phases for
  the phase diagram of the chiral ferromagnet MnSi.
\end{abstract}

\maketitle

\section{Introduction}

In metallic magnets without inversion symmetry -- so-called chiral
ferromagnets -- spin-orbit coupling effects determine the physics
decisively: they can twist the magnetization into a helix in the
ordered phase of the material. Prominent examples are MnSi and FeGe,
which have for a long time been known to exhibit spiral order
\cite{bak}. Recently, interest in such materials has been renewed with
a number of experiments that on the one hand found non-Fermi liquid
behaviour in a large temperature and pressure region of the phase
diagram \cite{pfleiderer}, and on the other hand uncovered signs of a
peculiar partially ordered state in neutron scattering experiments
\cite{nature}.

In the ordered state of MnSi, helical order is observed in neutron
scattering experiments in the form of Bragg peaks situated on a sphere
in reciprocal space. The radius of this sphere is proportional to the
inverse pitch of the helix, and the peaks are positioned in the
$\langle 1 1 1 \rangle$-directions. This locking of the helices is due
to higher order spin-orbit coupling effects. For temperatures down to
12 K the phase diagram exhibits a second-order phase transition out of
the helical into the disordered, isotropic phase when external
pressure is applied.  Below 12 K, helical order is lost at external
pressure of 12-14 kbar via a first-order phase transition into a
``partially-ordered'' state.  This state seems to retain remnants of
helical order on intermediate time and length scales: neutron
scattering experiments reveal a signal on the surface of a sphere in
reciprocal space, which is resolution-limited in the radial direction
but smeared out over the surface of the sphere \cite{nature}. Maxima
of the signal now point into the $\langle 1 1 0 \rangle$-directions
but are no longer sharply peaked.

Motivated by these experiments, we investigated ways of destabilizing
helical order in chiral ferromagnets and the possible new phases that can
exist besides helical order. Such phases are well known for
cholesteric liquid crystals\cite{bluePhases}, i.e.~chiral nematics,
where several so-called blue phases have been observed. In these
phases (more precisely in the phases I and II) crystalline order is
formed which can be interpreted as a periodic network of ordered
cylinders, see below. The lattice spacing is often of the order of a
few hundred nanometers, leading to the colorful appearance of these
phases (including the color blue in some variants).

As the Ginzburg-Landau theory for the director of a chiral nematic
liquid is very similar to that for the vector order parameter of a
chiral magnet (see discussion below), the question arises whether
similar phases are realized in magnetic systems. This question has
been addressed in some detail by Wright and Mermin\cite{bluePhases}
many years ago. They showed that amplitude fluctuations, which are
essential for the stability of blue phases, cost a factor $3$ more
energy for ferromagnets compared to liquid crystals and concluded that
such phases do not appear within a Ginzburg Landau theory with local
interactions.

Motivated by the physics of MnSi, several groups have reexamined this
issue.  R{\"o}\ss ler et al.~\cite{bogdanov} added a further parameter
to the Ginzburg Landau theory, allowing them to reduce the energy of
amplitude fluctuations arguing that such a term might arise from
higher-order fluctuation corrections. The term considered in
Ref.~[\onlinecite{bogdanov}] is, however, non-analytic in the
Ginzburg-Landau order parameter. In the presence of such a term, they
were able to show that a crystalline array of cylinders can have lower
energy than a uniform helix. We will discuss a similar structure
below. An alternative route (followed also by us) to destabilize the
helical solution are non-local interactions as suggested by Binz {\it
  et al.}\cite{binz}. They considered crystals formed from
superpositions of helices to be responsible for the partially ordered
phase of MnSi.

As pointed out by Wright and Mermin\cite{bluePhases}, is is useful to
distinguish two rather different limits when investigating the physics
of blue phases. Here one has to compare the correlation length $\xi$
for amplitude fluctuations of the ordered state (i.e. the width of a
typical domain wall) with the wavelength $2 \pi/q_0$ of the helical
state. For $\xi q_0 \gg 1$, the high-chirality limit, the magnetic
structure can be described by a superposition of a few helices. This
approach was taken by Binz {\it et al.}\cite{binz}.

The factor $\xi q_0$ is proportional to the strength of spin-orbit
coupling\cite{bak} and therefore expected to become large only very
close to the second-order phase transition where $\xi$ diverges. As
the pitch of the helix\cite{nature} in MnSi, $2 \pi/q_0\approx 175
\text{\AA}$, is very large compared to all other microscopic length
scales, we think that the low-chirality limit, $\xi q_0 \ll 1$, is
more appropriate for the description of the high-pressure phase of
this material. In this limit, amplitude fluctuations of the order
parameter cost more energy than twists of the phase and therefore one
has to look for order-parameter configurations with as little
amplitude fluctuations as possible.

In the following we will first investigate which terms in the Ginzburg
Landau theory destabilize the helical state. Then we will suggest
variational solutions appropriate in the low-chirality limit. Finally
we investigate experimental consequences of the resulting structures.

\section{Chiral magnets and blue phases}
Starting point of our investigation is a Ginzburg-Landau theory at
finite temperature, assuming that all modes with non-zero Matsubara
frequencies are massive and can be integrated out.

Up to second order in spin-orbit coupling, the Ginzburg-Landau theory
for chiral ferromagnets is given by \cite{bak}
\begin{equation} \label{fedensitysimple} f(\vec r) = \frac \a 2 \sum ({\bf \boldsymbol \nabla} M_i)^2 + \g \vec M
  \cdot ({\bf \boldsymbol \nabla} \times \vec M)+f_{\rm FM} ,
\end{equation}
where $\vec M = \vec M(\vec r)$ is the position-dependent
magnetization and  $f_{\rm FM}=\frac \d 2 (\vec
  M)^2+ u (\vec M)^4$ is the Landau free energy of the
  underlying ferromagnet. Spin-orbit coupling is present in the form of the
Dzyaloshinsky-Moriya interaction $\propto \g$. In order to facilitate
the following discussion, we also provide a second form for
(\ref{fedensitysimple}). It can be rewritten by setting $\vec M = \lambda \hat
n$, where $\lambda$ is the amplitude and $\hat n$ is the direction of the
magnetization $\vec M$:
\begin{multline} \label{enferro} f(\vec r) = \frac \a 2 \lambda^2
  \left(\nabla_i n_j + \frac \g \a \varepsilon_{ijk} n_k\right)^2 +
  \frac \a 2 ({\bf \boldsymbol \nabla} \lambda)^2 - \frac {\g^2}{\a} \lambda^2\\+f_{\rm FM}(\lambda).
\end{multline}
with $f_{\rm FM}(\lambda)=\frac \d 2 \lambda^2+u \lambda^4$.

Terms that break
rotational symmetry are of higher order
in spin-orbit coupling and take the form e.g.
\begin{multline} \label{anisotropic} B_1 \left[(\partial_x
    M_x)^2 + (\partial_y M_y)^2 + (\partial_z M_z)^2 \right] \\ + B_2
\left[ (\partial_x^2 \vec M)^2 + (\partial_y^2 \vec M)^2 +
  (\partial_z^2 \vec M)^2 \right] \\ + B_3
(M_x^4 + M_y^4 + M_z^4).
\end{multline}
(to fourth order in spin-orbit coupling) for the point group P2$_1$3
of MnSi \cite{helimagnons}. These terms will be neglected for the
moment. 

The helical state
\begin{equation} \label{helix}
\hat{\vec n}^{\rm helix} (\vec r) = \hat{\vec x} \cos(q_0 z) + \hat{\vec y} \sin(q_0 z),
\end{equation}
with $q_0 = \g/\a$ can be shown to be the lowest energy state of
(\ref{fedensitysimple}) if one assumes that the amplitude $\lambda$ of the
magnetization is constant and was argued in
Ref.~[\onlinecite{bluePhases}] to be the only ordered phase possible for
chiral ferromagnets.
Higher order terms in the Ginzburg-Landau expansion can, however,
serve to destabilize this helical order and induce other phases.  The
simplest (i.e.~of lowest order in spin-orbit coupling) of these terms
will be the focus of our investigation in this paper:
\begin{equation} \label{myterm}
 \left(\sum_i  M_i {\bf \boldsymbol \nabla}  M_i\right)^2 = \frac 14 (
 \boldsymbol \nabla \vec M^2)^2.
\end{equation}
This term acts only on the amplitude of the magnetization, not on its
direction. Therefore, it gives no contribution to the free energy
density in the helical phase, which has a uniform amplitude.  If
(\ref{myterm}) has a negative prefactor, then in a certain parameter
regime, it can be expected to destabilize helical order in favour of
an order parameter with a fluctuating amplitude. In the following we
will use the shorthand $(\vec M {\bf \boldsymbol \nabla} \vec M)^2$
for the expression (\ref{myterm}).

In order to determine the new states that could be stabilized by
(\ref{myterm}), it is instructive to explore the close analogies between
chiral ferromagnets and chiral liquid crystals \cite{bluePhases}. The
order parameter of chiral liquid crystals is a director, and as a
consequence topological defects are fundamentally different in both
systems. However, chiral liquid crystals can be described by a free
energy density that is quite similar in form to (\ref{enferro}), the
only differences being an additional cubic term in $\lambda$ and an extra
factor $1/3$ in front of the term $\propto (\boldsymbol \nabla \lambda)^2$. 

\begin{figure}
\includegraphics*[width=0.6 \linewidth]{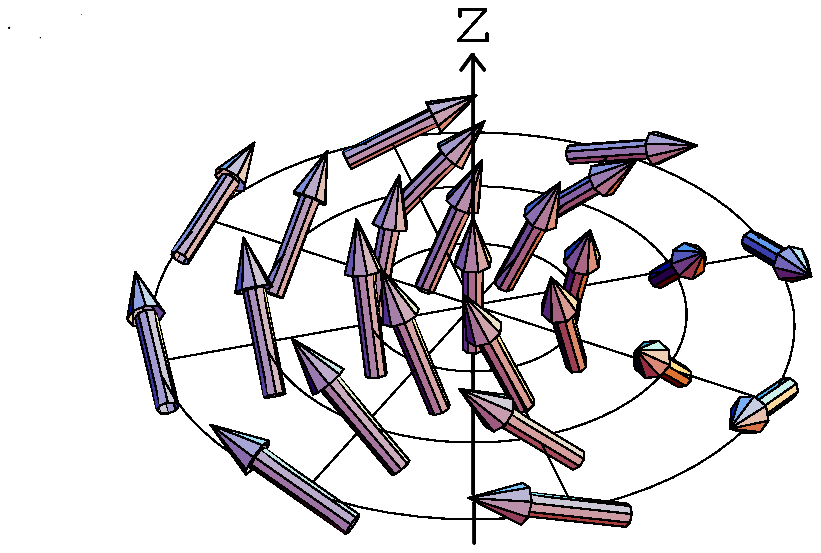}
\caption{\label{doubleTwist} Cut through a double-twist cylinder: 
  ``Double-twist''-configuration of the magnetization. Sheets of
  constant magnetization are wrapped as cylinders around a common
  axis.}
\end{figure}

{\em Locally}, the blue
phases are based on a configuration of the magnetization
that can be shown to be even lower in energy than the helix:
\begin{equation} \label{doubletwist}
\hat{\vec n}^{\rm dt} (\vec r) = \hat {\vec z} \cos(q r) - \hat{\vec
  \phi} \sin(q r),
\end{equation}
in cylinder coordinates, see Fig.~\ref{doubleTwist}.  This can easily
be seen\cite{bluePhases} by comparing the energy density of the
uniform helix, $-\frac{\gamma^2}{2 \alpha} \lambda^2+f_{\rm
  FM}(\lambda)$, to the result obtained by plugging
Eq.~(\ref{doubletwist}) into Eq. (\ref{enferro}): For
$\lambda=const.$ and $q=q_0$, the first two terms in (\ref{enferro}) vanish for
$r=0$, and therefore the energy density at $r=0$ is given by
$-\frac{\gamma^2}{ \alpha} \lambda^2+f_{\rm FM}(\lambda)$, i.e.
lowered by an extra factor $-\frac{\gamma^2}{2 \alpha} \lambda^2$.

This so-called ``double-twist'' configuration is
cylindrically symmetric: sheets of constant magnetization are rolled
up around a common cylinder axis (see Fig.~\ref{doubleTwist}). 
The configuration (\ref{doubletwist}) is, however, only favoured in
the vicinity of the cylinder axis -- the energy difference between the
helix and the double-twist configuration diminishes and even becomes
positive as the distance from the cylinder axis is increased.
Isolated double-twist configurations therefore necessarily have to
occur in the form of cylinders with an amplitude that becomes zero at
a certain distance from the cylinder axis. However, amplitude
fluctuations cost energy, as can be seen from the second term in
(\ref{enferro}). In the case of liquid crystals, the prefactor of
$({\bf \boldsymbol  \nabla} \lambda)^2$ is small enough for double-twist cylinders to become
energetically lower in energy than the helix (see discussion above).
Crystals made of these double twist cylinders are indeed believed to
lie at the heart of the blue phases I and II  in liquid crystals
\cite{bluePhases}. In the case of the ferromagnet, however, the
energy cost of amplitude fluctuations outweighs the energy gains due
to directional fluctuations within the
double-twist structure, and for
a free energy density given by (\ref{fedensitysimple}) (single)
double-twist cylinders do not occur \cite{bluePhases}.

Adding the term (\ref{myterm}) to the free energy density can
invalidate this conclusion: if the prefactor of (\ref{myterm}) is
allowed to become negative, it can reduce the cost of amplitude
fluctuations and allow for the appearance of blue phases even in
chiral ferromagnets. If (\ref{myterm}) has a negative prefactor, then
the inclusion of higher order terms in the Ginzburg-Landau theory
becomes necessary in order to obtain a stable solution and avoid
e.g.~an unbounded magnetization.

In the following, we want to analyze the possible occurrence of the
analogue of blue phases in chiral ferromagnets. By rescaling the
magnetization, the free energy density and the momenta, the free
energy density can be cast into the following
form:
\begin{multline} \label{myFED}
f(\vec r) = \d \vec M^2 + \sum ({\bf \boldsymbol \nabla}
    M_i)^2 + \vec M \cdot ({\bf \boldsymbol \nabla} \times \vec M) + \vec M^4 \\+
\xi (\vec M {\bf \boldsymbol \nabla} \vec M)^2 + \eta \, h_i (\vec M),
\end{multline}
where $\xi < 0, \ \eta > 0$ and $h_i (\vec M)$ is a term containing
$k$ powers of $\vec M$ and $l$ derivatives ($k > 4$, $l \ge 2$), which
is used to stabilize solutions against an unbounded magnetization and
oscillations thereof. Possible choices are for example $h_1(\vec M) =
(\vec M {\bf \boldsymbol \nabla} \vec M)^2 \sum ({\bf \boldsymbol
  \nabla} M_i)^2$, $h_2(\vec M) = (\vec M {\bf \boldsymbol \nabla}
\vec M)^4$ and $h_3(\vec M) = \vec M^2 (\vec M {\bf \boldsymbol
  \nabla} \vec M)^2$. In the rest of this paper, we will use $h_1(\vec
M)$ exclusively: this term is the one with the least number of
derivatives and powers of $\vec M$ that on the one hand stabilizes
single double twist cylinders and on the other hand is identically
zero for the single helix.

\section{From cylinders to crystals}
As a first step, we investigated single double twist cylinders for $\d
= 1/4$. At this point, the helical and isotropic phase are degenerate
with $f^{\rm helical} = f^{\rm isotropic} = 0$. If a single cylinder
can now be shown to have negative free energy, then the system can
lower its free energy further by creating extended networks of
double-twist cylinders: crystalline phases, the analogy to blue phases
in liquid crystals, can be expected to form in chiral ferromagnets.

\begin{figure}
\includegraphics*[width=0.9 \linewidth]{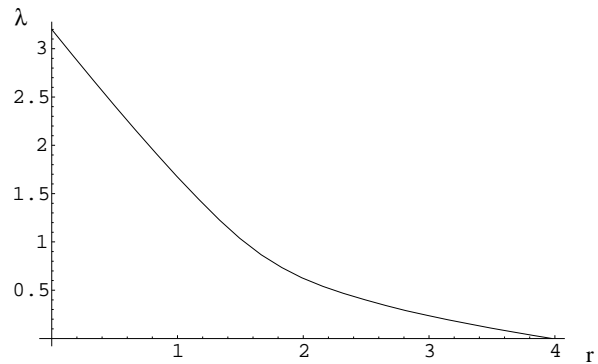} 
\caption{\label{amplitude} Amplitude function $\lambda(r)$ which
  minimizes (\ref{myFED}) using the directional 
dependence (\ref{doubletwist}), the
  stabilizing term $h_1$ (with $\xi = -6.5$, $\eta = 5$) and the
  boundary condition $\lambda(r)=0$ for $r>R$. Here $q=0.38 q_0$ and $R=3.98$ are free
variational parameters where $q_0=1/2$ is the helix wave vector. The resulting
 free energy gain per length is given by $f^{\rm cyl} = -0.21$.
}
\end{figure}

Setting $\vec M = \hat{\vec n}^{\rm dt} (\vec r) \lambda(r)$, with
$\hat{\vec n}^{\rm dt} (\vec r)$ given by (\ref{doubletwist}), we
calculated numerically the amplitude function $\lambda(r)$ that minimizes
the free energy density (\ref{myFED}) (for certain values of $\xi$ and
$\eta$), subject to the condition that the magnetization drops to zero
at a certain distance from the cylinder axis. For a given $\eta$ and
sufficiently negative values of $\xi$, single cylinders are stable
configurations within a Ginzburg-Landau theory of the form
(\ref{myFED}), see Fig.~\ref{amplitude}. In order to minimize its free
energy, the system will try to produce many such cylinders, packed as
tightly as possible.  In these configurations, the magnetization only
has to drop to zero on lines or points, if at all.  Considering that
it is the competition between phase and amplitude fluctuations that
either stabilizes double-twist structures or not, it can be expected
that crystalline arrangements of the helices can exist even in
parameter regimes where single cylinders are unstable.

\begin{figure}
\begin{tabular}{cccc}
(a) & \hspace{-4mm} \includegraphics*[width=0.45 \linewidth]{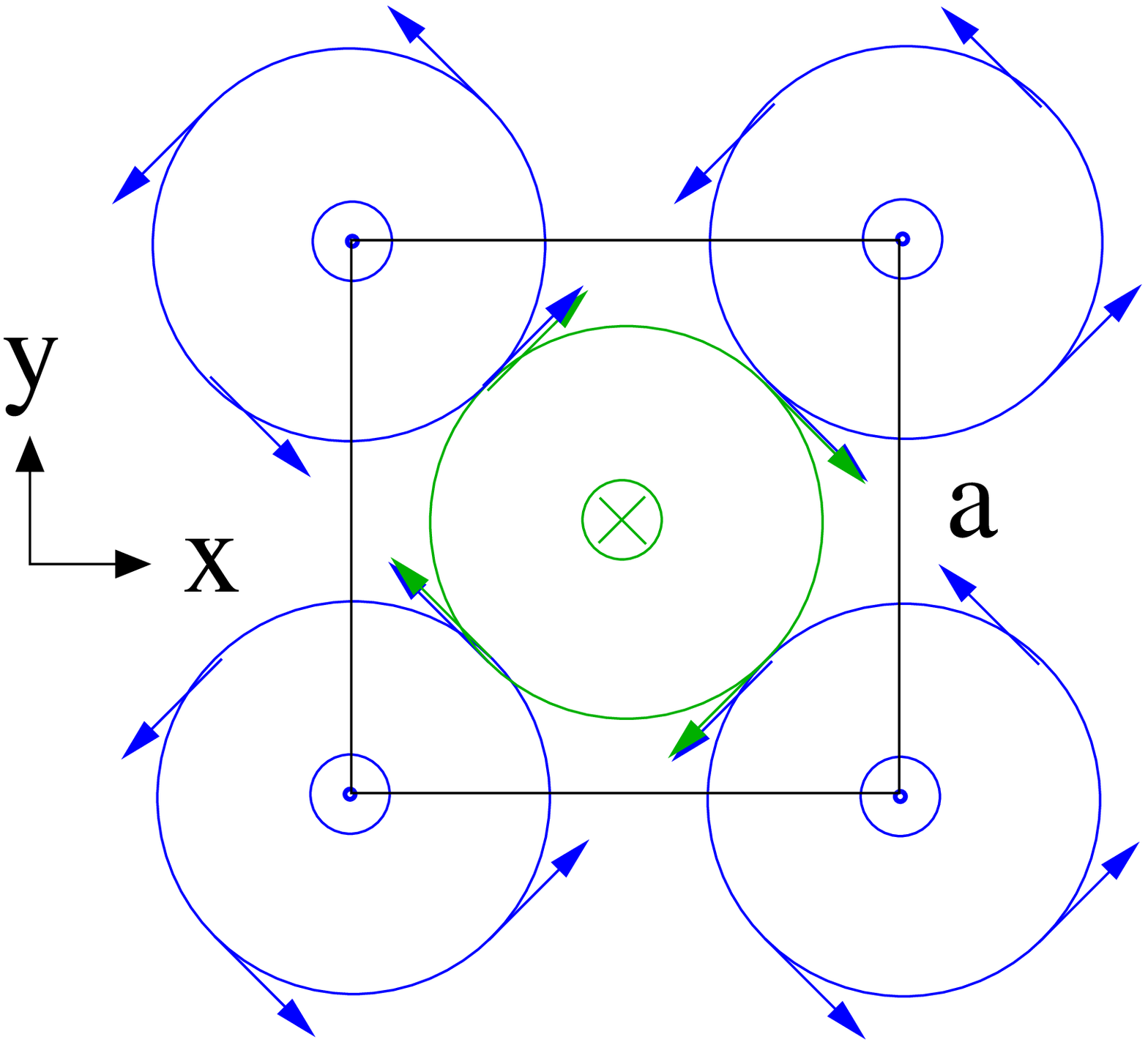}
& (b) & \hspace{-4mm}\includegraphics*[width=0.5 \linewidth]{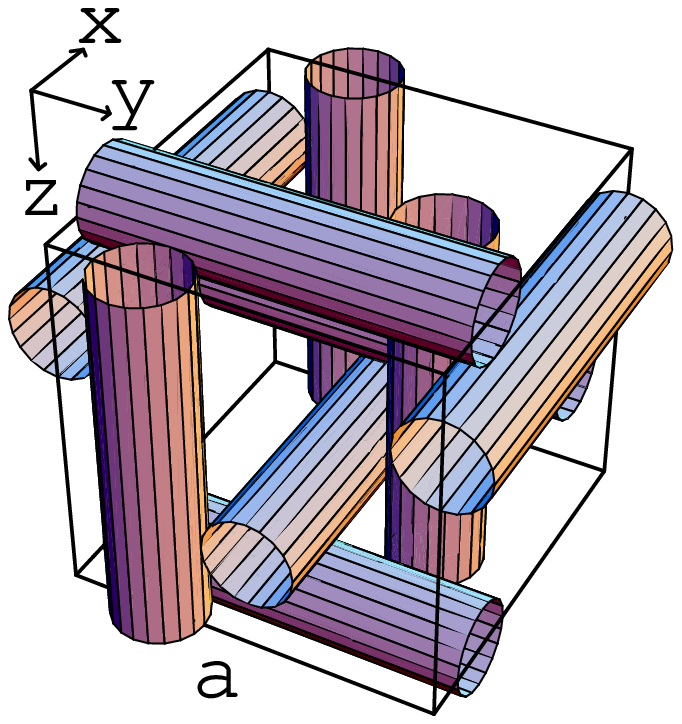}
\end{tabular}
\caption{\label{crystals} Crystalline structures built from
  double-twist cylinders: (a) Unit cell of the square lattice of
  double-twist cylinders in the x-y-plane. The magnetization on the
  cylinder axes as well as at the points where cylinders touch is
  shown explicitly. (b) Unit cell of the cubic lattice of double-twist
  cylinders. At the points where cylinders touch, the magnetization
  has twisted to an angle of 45$^\circ$ from the cylinder
  axis.}
\end{figure}

The possible crystalline structures
are subject to the condition that the magnetization matches where
cylinders touch, in order to avoid discontinuities in the
magnetization. This condition is much more restrictive for
ferromagnets compared to liquid crystals\cite{bluePhases}. We have
found\cite{footnote} only two allowed structures
constructed as networks of double-twist cylinders: a square and a cubic
lattice, see Fig.~\ref{crystals}. The square lattice is invariant with
respect to rotations by multiples of $\pi/2$ around the z-axis,
translations along the z-axis as well as a translation by $a(\frac 12,
\frac 12, 0)$ combined with time reversal ($\vec M \to - \vec M$).  In
this structure, the magnetization has to go to zero on the line
$(\frac a2,0,z)$ and symmetry-equivalent lines.  The symmetry
transformations that leave the second structure (cubic lattice)
invariant are cyclic permutations of the axes and a translation by
$a(\frac 12, \frac 12, \frac 12)$ combined with time reversal. In this
structure, the magnetization vanishes only at $a \left(\frac 18, \frac
  58, \frac 38\right)$, $a \left(\frac 38, \frac 38, \frac 38 \right)$, and
symmetry-equivalent points in the unit cell.

We calculated the free energy density of these structures by means of
a variational ansatz for the amplitude of a single double-twist
cylinder:
\begin{equation} \label{amplitudeAnsatz}
\lambda(r) = y_0 \cdot (r_0 - r) e^{-r/r_1} \Theta(r_0 - r),
\end{equation}
where $y_0$, $r_1$ and the double-twist wave vector $q$ are
variational parameters. This ansatz is based on the numerics for a
single cylinder. The cylinders were then arranged in crystals as shown
in Fig.~\ref{crystals}. The cutoff $r_0 \approx \sqrt 5 \pi /(2 q)$
was chosen for  convenience to avoid within the calculation the tiny overlap of
cylinders which are far apart.

\begin{figure}
\includegraphics*[width=0.9 \linewidth]{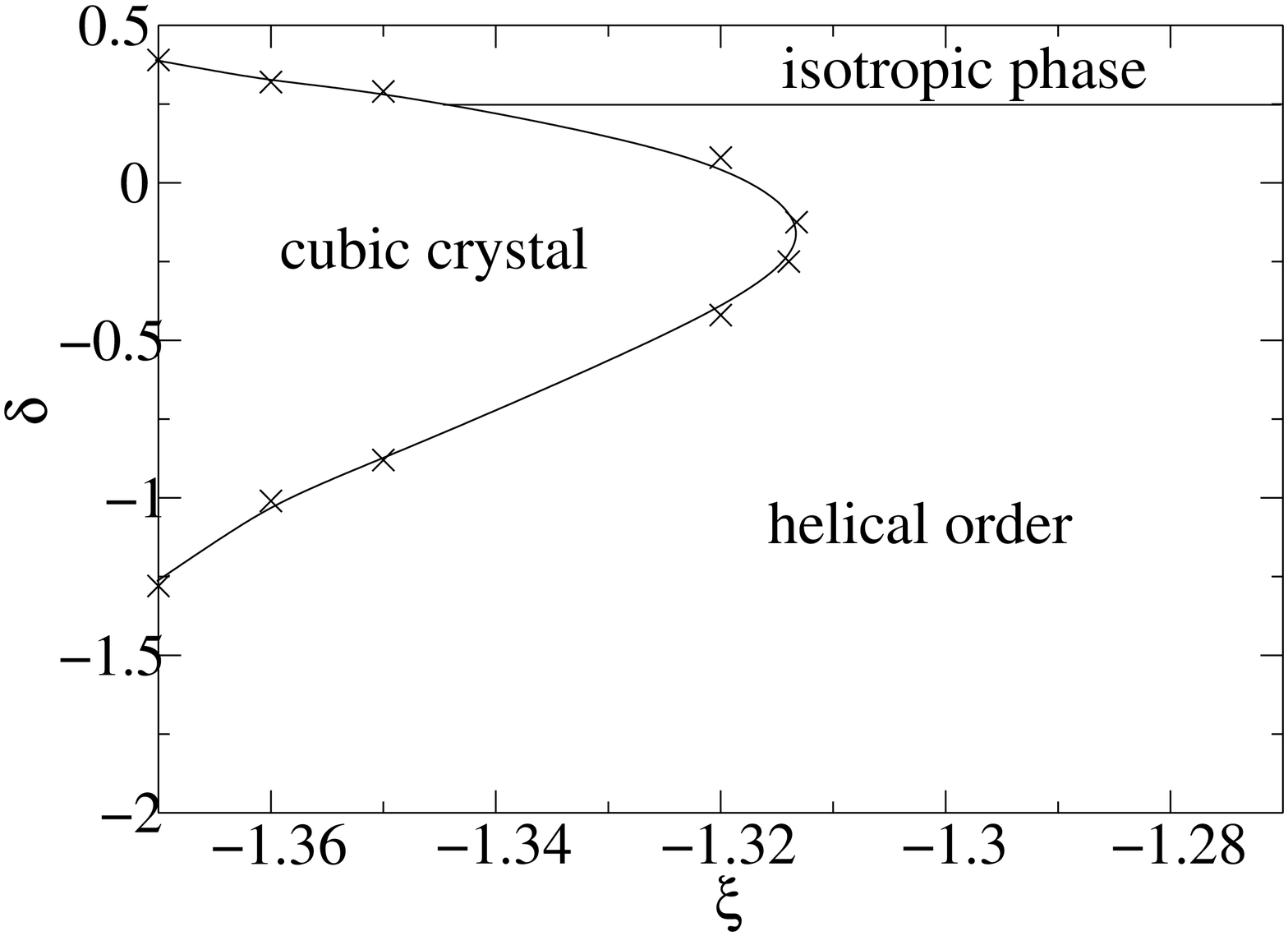} 
\caption{\label{phaseDia} Phase diagram for $\eta = 0.05$. In addition
  to the helical and the isotropic phases there is a parameter range
  where the cubic lattice of double-twist cylinders minimizes the free
  energy.}
\end{figure}

A phase diagram as a function of the remaining three free parameters
can now be computed. Here, we set $\eta$ to a fixed value, assuming
that it is the least susceptible to variations of external pressure,
and established the phase diagram as a function of the two remaining
free parameters. The result for $\eta = 0.05$ can be seen in
Fig.~\ref{phaseDia}. 

\begin{figure}
\includegraphics*[width=0.9 \linewidth]{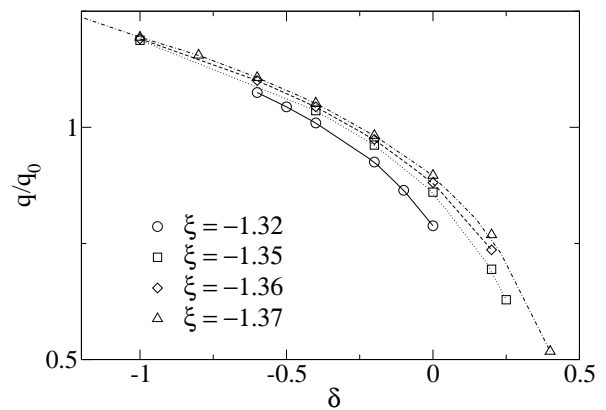}
\caption{\label{waveVector} Ratio of double twist wave vector $q$ [see
 Eq.~ (\ref{doubletwist})] versus helix wave vector $q_0$ [see
  Eq.~(\ref{helix})] for $\eta = 0.05$: $q/q_0$ varies considerably
  but remains of order
  1 in the parameter region where the cubic crystal is stable.}
\end{figure}

One immediately notices that there is no parameter regime in which the
square lattice shown in Fig.~\ref{crystals} (a) has the lowest free
energy density; this is also true for different values of $\eta$ not
shown here. While the filling fraction of double-twist regions is
certainly larger in the square lattice than in the cubic structure,
the magnetization in the cubic crystal only has to go to zero at
points and not on lines, as it is the case for the square lattice.
Furthermore, in the cubic lattice the magnetization only has to twist
outwards by 45 degrees until cylinders touch, compared to 90 degrees
in the case of the square lattice (see Fig.~\ref{crystals}). These
factors conspire to make the free energy of the cubic structure even
lower than that of the square lattice.  Within our variational ansatz
we find that the 
transition from the crystalline to the isotropic phase is very weakly first
order [the relevant free energy differences are less than $0.02$ in
units of Eq.~(\ref{myFED})]. Note, however, that our analysis might
break down close to the transition, as one enters the high-chirality regime,
see above.

Since the double twist wave vector $q$ is now also a variational
parameter, it is no longer necessarily identical to the helix wave
vector $q_0$. In Fig.~\ref{waveVector} we show that $q$ depends
considerably on the microscopic parameters but remains of order $q_0$.

What signal can these structures be expected to produce in neutron
scattering experiments? As for any crystal, peaks in neutron
scattering originate from the lattice structure itself, and the
configuration of the magnetization within the Wigner-Seitz cell only
enters in the shape of form factors. The lattice constants are 
functions of the double-twist wave vector and determined by the
requirement that the magnetization has to match where cylinders touch.
For the square lattice, one obtains $a = \sqrt{2} \pi/q$, and for the
cubic lattice $a = 2 \pi/q$. 

In neutron scattering, higher order Bragg peaks can also be expected
to be be generated by the lattice structures. The
elastic scattering crossection\cite{neutrons} can be calculated from a
Fourier transform of the magnetization:
\begin{equation}
\left(\frac{d\s}{d\W}\right)_{el} \propto |\hat{\vec \kappa} \times (\vec M(\vec \kappa)
  \times \hat{\vec\kappa})|^2 . 
\end{equation}
In order to compare with experimental data, the quantities of interest
are the positions of the Bragg peaks and the relative intensities of the
 Bragg peaks.

For the square lattice, normalizing intensities to give unity for the
first reflection with Miller indices $\langle 1 0\rangle$, the $\langle 2 1\rangle$
and the $\langle 3
0\rangle$-reflections have intensities 0.08 and 0.01, respectively. The
invariance of the square lattice with respect to a translation by $a (
\frac 12, \frac 12)$ combined with $\vec M \to - \vec M$ constrains
all Bragg peaks $\langle h k \rangle$ with $h+k=2 n$ to vanish. In the
case of the square lattice, the lowest order reflexes already account
for 84\% of the total intensity.

For the cubic lattice, assuming that the lowest Bragg peak, i.e.~the
$\langle 1 0 0 \rangle$-peak, has intensity 1.0, the $\langle 2 1
0\rangle$- and $\langle 3 0 0\rangle$-peaks have relative intensities
0.17 and 0.03. The $\langle 1 1 0\rangle$-, and $\langle 2 0
0\rangle$-peaks vanish for symmetry reasons, while the $\langle 1 1
1\rangle$-peak vanishes as a consequence of our ansatz based on a
linear combination of cylinders in $x$, $y$ and $z$ direction, and
should really be non-zero as allowed by symmetry. Our approach
suggests, however, that these peaks have small weight. The $\langle 1
0 0 \rangle$-peaks of the cubic lattice only represent 46\% of the
total scattering intensity $I_{\rm cubic}$. The $\langle 1 0 0
\rangle$-, $\langle 2 1 0 \rangle$- and $\langle 3 0 0 \rangle$-peaks
of the cubic lattice add up to 79\% of $I_{\rm cubic}$.

Anisotropic terms that orient the helix also act on the crystalline
structures and determine their orientation with respect to the crystal
lattice of the substance. For the square lattice, where all cylinders
are arranged in parallel, we find using Eq.~(\ref{anisotropic}) for
$B_1 = B_2 = 0$, $B_3 \neq 0$ that weak anisotropic terms align the
cylinder axes parallel to the preferred direction of the helix vector,
i.e. either in the $\langle 1 1 1 \rangle$ or $\langle 1 0 0 \rangle$
direction, depending on the sign of $B_3$. However, the orientation of
the crystal in perpendicular direction is not affected to leading
order by the anisotropy term, but higher-order terms would lock a
perfect crystal.  Motivated by the ``partial order'' observed in the
high-pressure phase of MnSi (see introduction), we investigate the
expected signature in neutron scattering assuming that these
higher-order terms are not effective. In this case, the square lattice
will produce rings in planes normal to the $\langle 1 1 1
\rangle$-direction (the orientation of the helix in the low-pressure
phase). These rings intersect to produce maxima in the $\langle 1 1 0
\rangle$-direction on a circle with radius $\sqrt 2 q$ in reciprocal
space. For parameters with $\sqrt 2 q \approx q_0$, this is consistent
with the observed signatures\cite{nature} in neutron scattering. Note,
however, that at least within our model, the square lattice never has
the lowest energy.

In Ref.~[\onlinecite{bogdanov}] , it was argued that an amorphous
texture of parallel cylinders (which the authors called skyrmions)
aligned preferentially along the $\langle 111 \rangle$ direction would
also produces such rings. However, such a scenario would not explain
the resolution limited width in radial direction, i.e. at least on
length scales of 2000\AA\ the square crystal would have to remain
intact.

In the case of the cubic lattice, the numerical calculation of the
free energy density shows that for $B_3>0$ the anisotropic terms are
minimized if one of the axes of the cubic lattice is $\| (1/\sqrt 2,
1/\sqrt 2, 0)$ and the other axes are $\| (-1/2, 1/2, 1/\sqrt 2)$ and
$\| (1/2, -1/2, 1/\sqrt 2)$, respectively. While such structures would
produce peaks in the $\langle 1 1 0\rangle$ direction (as observed in
the high-pressure phase of MnSi), one expects also considerable
intensity rather close to $\langle 1 1 1\rangle$ (as e.g.~ $(-1/2,
1/2, 1/\sqrt 2)$ differs only by about 10 degrees from $(-1 1 1)$).
Experimentally, however, one observes a minimum of the intensity in
the $\langle 1 1 1\rangle$ direction. For $B_1 = B_2 = 0$, $B_3 < 0$,
the system would prefer to align the cubic structure with the crystal
lattice of MnSi.

Binz et al.~\cite{binz} have argued that the Bragg peaks in neutron
scattering should be smeared out because the magnetization has a
varying amplitude and is therefore susceptible to interactions with
non-magnetic impurities. The same argument would apply for the
crystalline structures presented in this paper.

\section{Conclusions}
In conclusion, we have constructed ``blue phases'' in chiral
ferromagnets. Is it likely that these blue phases are realized in
MnSi? The signatures in neutron scattering seem to be consistent with
the square structure (also considered in
Ref.~[\onlinecite{bogdanov}]), which is, however, never the
ground state within the models which we considered. Our results suggest
that the intensity in neutron scattering is located on rings
intersecting in the $\langle 110 \rangle$-direction. It would be
interesting to check experimentally whether the intensity distribution
arising from such a picture fits quantitatively.

In the case of the cubic lattice, the positions of the expected maxima
do not match the ones observed. In any case, the smoking gun
experiment to detect crystalline structures, would be the observation
of higher order Bragg peaks (this applies also for other structures,
e.g. those suggested in Refs.~[\onlinecite{bogdanov,binz}]) and we
hope that our estimates may guide future experiments in this
direction.

However, it is still unclear whether the partially ordered state of
MnSi is indeed a separate phase. Recent measurements of the thermal
expansion coefficient give no trace of a phase transition from the
partially ordered state to the isotropic phase\cite{privateChris}.
Furthermore, $\mu$SR experiments\cite{muSR} give no evidence for
static order in this regime and it remains unclear on which time
scales the partial order survives\cite{nature}.

It therefore remains an open question whether distinct phases other
than the helical one are present in chiral ferromagnets that are
currently investigated experimentally. In fact, the partially ordered
phase in MnSi might be more analogous to the blue phase III than to
blue phases I and II. While the precise structure of the blue phase
III is not completely clear, it seems to show some type of crystalline
short range order but is a liquid on long length
scales\cite{chiralityBuch,citeFromNature}. Therefore it has been
conjectured \cite{onecylinder} that an amorphous arrangement of
double-twist cylinders is relevant for this blue phase III in liquid
crystals -- this might also be
true\cite{nature,blueFog,bogdanov} for the partially
ordered phase in MnSi.

\acknowledgments
We thank B. Binz, C. Pfleiderer, U. K. R\"o\ss ler, A. Vishwanath for
useful discussions and the SFB 608 of the DFG for financial support.


\begin{thebibliography}{99}
\bibitem{footnote}Although the three-dimensional structures proposed in
Ref.~[\onlinecite{binz}] cannot be realized with double-twist
cylinders, they nonetheless contain large, clearly identifyable
regions that exhibit double-twist-structures.
\end{thebibliography}

\end{document}